\documentclass[aps,prb,10pt,twocolumn,floatfix,showpacs,superscriptaddress,longbibliography]{revtex4-1}

\usepackage{graphicx,amsmath,amsfonts,bm}
\usepackage[colorlinks=true, citecolor=blue, linkcolor=blue, urlcolor=blue]{hyperref}
\usepackage[all]{hypcap}
\usepackage{bbold}
\usepackage{ulem}
\usepackage{soul}
\usepackage{xcolor}

\usepackage{comment}

\graphicspath{{./Fig/}}

\begin{document}
	
	\title{Rashba-controlled thermal valve in helical liquids}
	\author{Alessio Calzona}
	\affiliation{Institute of Theoretical Physics and Astrophysics, University of W\"urzburg, 97074 W\"urzburg, Germany}
\author{Niccol\`o Traverso Ziani}
\affiliation{Dipartimento di Fisica, Universit\`a di Genova, Via Dodecaneso 33, 16146, Genova, Italy}
\affiliation{CNR-SPIN, Via Dodecaneso 33, 16146, Genova, Italy}
\author{Matteo Carrega}
\affiliation{CNR-SPIN, Via Dodecaneso 33, 16146, Genova, Italy}
\author{Maura Sassetti}
\affiliation{Dipartimento di Fisica, Universit\`a di Genova, Via Dodecaneso 33, 16146, Genova, Italy}
\affiliation{CNR-SPIN, Via Dodecaneso 33, 16146, Genova, Italy}

	\begin{abstract}
In the context of one-dimensional fermionic systems, helical Luttiger liquids are not only characterized by intriguing spin properties, but also by the possibility to be manipulated by means of electrostatic gates, exploiting finite Rashba coupling. We use this property to show that a heterostructure composed of a helical Luttinger liquid, {contacted to two metallic leads and supplemented by top gates,} can be used as a tunable thermal valve. By relying on bosonization techniques and scattering of plasmonic modes, we investigate the performance of this valve with respect to electron-electron interactions, temperature, and properties of the gates. The maximal modulation of the thermal conductance that the proposed device can achieve is, for experimentally relevant parameters, around $7 \%$. Such variation can be both positive or negative. Moreover, a modification in the geometry of the gate can lead to particular temperature dependencies related to interference effects. {\color{black}We also argue that the effects we predict can be used to establish the helical nature of the edge states in two-dimensional topological insulators.}
	\end{abstract}

\maketitle

\section{Introduction}
Interacting one-dimensional (1D) electronic channels have attracted great theoretical and experimental interest during the last decades. Due to the reduced spatial dimensionality, Coulomb interactions play a prominent role\cite{giamarchi2003quantum, deshpande2010electron, barak2010interacting}, resulting in a non-Fermi liquid behaviour and in peculiar effects such as charge fractionalization or spin-charge separation\cite{safi1995transport,sass, Auslaender01042005, lorenz2002evidence, bockrath1999luttinger, steinberg2007charge, kamata2014fractionalized, calzona1,esa,d1}. The scenario is even richer when spin-orbit coupling (SOC) is present\cite{qi11, hasan10, bernevig06, pradareview}. In particular, the so-called helical liquids can emerge. Such states are described in terms of a couple of 1D channels whose propagation direction and spin degree of freedom are tightly bound. This property, called spin momentum locking, has been predicted and experimentally verified in several and diverse systems. Among them, the edge states of two-dimensional topological insulators based on HgTe/CdTe heterostructures\cite{qi11, hasan10, konig07, d2}, strong SOC nanowires\cite{picciotto10, Knez14, Du15,n1}, Bismutene flakes\cite{stuhler20}, and high quality graphene samples with large dielectric substrate and enhanced SOC contributions\cite{veyrat}. As in most 1D systems, interaction effects are visible even in helical liquids. Fingerprints related to electronic correlations have been already reported by means of transport\cite{Du15, strunz} and spectroscopic measurements\cite{stuhler20}.

A specific and promising property of the systems just mentioned is that the strong SOC allows for the manipulation of a Rashba-like contribution\cite{dolcini1, dolcini2, ronetti20, privitera2020}. To this end, external electric fields, induced for example by properly patterned side or top gates, can be used. The possibility to locally modify the SOC component by electrical means can be of interest both for fundamental studies and for possible applications for quantum technological purposes, in particular in spintronics\cite{Michetti_APL13,Linder_NPhys15, Breunig_PRL18}.

In this context, a recently emerging field of research revolves around the exploitation of thermal gradients, instead of voltage drops, in nanodevices. The heat and energy flows in nanostructures\cite{Pekola15, Vinjanampathy16, Benenti17, Sanchez2014, Pekola2000,Miller2006,Giazotto2006,Chowdhury2009,d3,Vischi2018} have hence been addressed. Interesting results have been already obtained within different platforms (using both normal and superconducting nanostructures), demonstrating the coherent control and manipulation of heat flux\cite{Pekola15, Giazotto2006, Fornieri2017}. These realizations are often based on hybrid systems with diffusive transport properties. However, measurements in the quantum point contact geometry in ballistic channels driven by thermal gradients have also been reported\cite{Pekola15, larocque2020, duprez2021}. The study of the transport properties characterizing helical liquids in presence of thermal gradients has recently been carried out as well, focusing mostly on non-interacting systems and on Josephson-like configurations with superconducting leads\cite{bours2018, ronetti2017, blasi2020, scharf2021}. In the absence of SOC, on the other hand, the violation of the Wiedemann-Franz (WF) law, as expected for a non-Fermi liquid system, has been predicted\cite{kane, fazio, filippone, krive,garg2009} and reported\cite{grossviolation} in 1D systems.

{In this work, we consider at the same time the effects of Rashba coupling and electron-electron interactions. In particular, we show how the Rashba contributions can be used to manipulate the transport properties of an interacting helical liquid in presence of a thermal gradient.} We mostly focus on the energy flow. Specifically, we consider an inhomogeneous helical Luttinger liquid (HLL)\cite{calzona1, dolcini1, muller}, subject to a thermal gradient, where one or more capacitively coupled top gates can induce variations in both the interaction strength (via screening effect) and the Rashba coupling.

Focusing on a thermally-biased two-terminal configuration, we show that the thermal conductance can be manipulated by varying the Rashba coupling strength. This effect can be exploited to engineer a thermal valve, and therefore to selectively suppress (or enhance) the energy flow. We characterize this behavior by evaluating the efficiency of the thermal valve in different configurations. In particular, we discuss how the performance of the thermal valve can be affected by different covering ratios of a top gate with respect to the length of the interacting helical liquid. We show that the choice of the best configuration depends on the interaction strength, which also influences the tunability of the gate-induced Rashba effect. Interestingly, we demonstrate that the performance of the thermal valve can be further improved in a configuration with more than one top gate.

It is worth mentioning that other platforms, involving superconducting elements, have been recently proposed as efficient thermal valves\cite{Fornieri2017, strambini1, strambini2, ronzani2018}. These systems show great performances at cryogenic temperatures (well below the critical temperature of the inherent superconducting compounds). However, the associated efficiencies are found to be comparable with the ones described in this work, where the working principle can be used also at higher temperature since it is not based on superconducting correlations.

The rest of the article is organized as follows. In Sec.\ref{sec:model}, we present the model and the solution of the equations of motion for an inhomogeneous helical system. In Sec.\ref{sec:results}, we discuss the results, focusing on the transport properties in presence of a thermal bias in a two-terminal configuration. We argue that a gate tunable thermal valve can be designed by exploiting the properties of an interacting helical liquid. {\color{black}We also reverse the argument and show that a thermal transport experiment can be useful in determining if the edge states are helical.} Section \ref{sec:summary} is devoted to a summary of our main results. Technical details can be found in the appendices.

\section{Model}
\label{sec:model}
\subsection{General setting}
\begin{figure}
 	\centering
 	\includegraphics[width=0.95\linewidth]{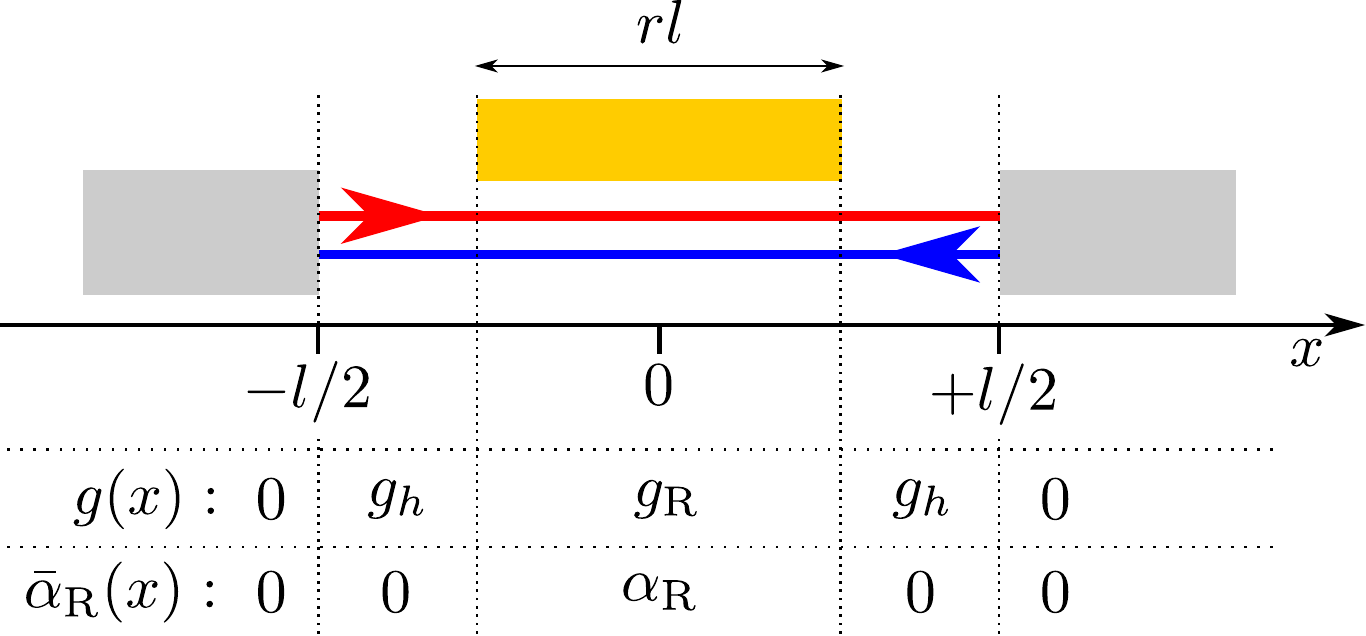}
 	\caption{Sketch of the setup. The helical system, consisting of spin up 
(red) and spin down (blue) counterpropagating channels, is contacted with 
two leads (grey). A top gate (yellow) covers a fraction $r$ of the helical system and can induce, and therefore can tune, a finite Rashba interaction.}
 	\label{fig:setup}
 \end{figure}

We consider a inhomogeneous helical system with spatially varying electron-electron interactions. The helical liquid consists of two one-dimensional counterpropagating electronic channels with opposite spin polarization, described in terms of the spinor $\Psi(x) = 
(\psi_\uparrow(x), \; \psi_\downarrow(x))^T$. At low energies the spectrum is linear{\cite{qi11, hasan10, dolcetto}} and the Hamiltonian density can be written as ${\cal H}_{{\rm hll}}(x)={\cal H}_0 (x)+ {\cal H}_{{\rm int}}(x)$, with the free contribution (hereafter 
we set $\hbar=k_{{\rm B}}=1$)  
\begin{equation}
\label{eq:h0}
	\mathcal{H}_0(x) = -i v_{\rm F} \Psi^\dagger(x) \partial_x \sigma_3 \Psi(x) 
\end{equation}
where $v_{\rm F}$ is the Fermi velocity and $\sigma_i$ ($i=1,2,3)$ are Pauli matrices acting on the spin degree of freedom. The second contribution is related to electron-electron (e-e) interactions. Assuming short range correlations{\cite{giamarchi2003quantum, barak2010interacting, dolcetto}} it can be written in terms of density-density operators and it reads
\begin{equation}
\label{eq:h1}
	\mathcal{H}_{\rm int}(x) = g(x) \rho_\uparrow(x) \rho_\downarrow(x) 
\end{equation}
where $\rho_\sigma (x) = \psi_\sigma^\dagger(x) \psi_\sigma(x)$ is the electronic density with spin index $\sigma  = \uparrow, \downarrow$.
This term describes inter-channel interactions, determined by the spatially-dependent coupling $g(x)$\cite{dolcini1, calzona1, muller}. We notice that in the expression above we have neglected intra-channel density-density interactions $\propto\rho_\sigma(x)\rho_\sigma(x)$ since the corresponding coupling strength is usually vanishingly small\cite{gi1,gi2}. We underline however that inclusion of (small) finite intra-channel interactions does not qualitatively affect the 
results discussed below, resulting solely in a quantitative renormalization of the propagation velocity\cite{dolcini1, strunz}. The parameter $g(x)$ entering Eq.\eqref{eq:h1} describes the coupling strength of e-e interactions and it is assumed space-dependent. As sketched in Fig.\ \ref{fig:setup}, we consider three different regions. Two metallic leads (depicted in grey) extending for $x<-l/2$ and $x>l/2$. Physically, they can model two non-interacting electron reservoirs\cite{safi1995transport, calzona1} connected to the central helical system{\cite{safi1995transport, calzona1, muller}}. These reservoirs can be effectively modelled as one dimensional non-interacting channels described by Eq.~\eqref{eq:h0} with $g(x)=0$.
 Conversely, the helical liquid, located within $-l/2 \leq x\leq l/2$, exhibit a finite e-e interactions, modeled by $0<g(x)<1$.\\

 Moreover, in between the two leads, a centered top metallic gate (depicted in yellow in Fig.~\ref{fig:setup}) covers a fraction $r$ of the helical system and allows for the local control of the electric field. The latter generates a space-dependent voltage-induced Rashba SOC $\bar \alpha_{{\rm R}}(x) \geq 0$ \cite{dolcini1,Vayrynen2011,Woljcik2014} which results in the Hamiltonian density \cite{dolcini1, dolcini2}
 \begin{equation}
	\mathcal{H}_{\rm R}(x) = - \frac{i}{2} \Psi^\dagger(x)\, \{\bar \alpha_{{\rm R}}(x),\partial_x \}\,\sigma_2 \Psi(x). 
\end{equation}
In particular, we consider a uniform and tunable $\bar \alpha_{{\rm R}}(x)=\alpha_{{\rm R}}\geq0$ under the top gate and $\bar \alpha_{{\rm R}}(x)=0$ elsewhere (see Fig.\ \ref{fig:setup}). Moreover, the presence of a top gate can result in possible screening effect, that can lower the strength of e-e interactions. We will take into account this possibility, by considering two different values of the e-e interaction strength. As sketched in Fig.\ \ref{fig:setup}, we consider a coupling constant $g = g_h \geq 0$ for the regions which are not covered by 
the gate and a (possibly) smaller $0\leq g_{{\rm R}} \leq g_h $ for the regions right under the gate. 
\subsection{Inhomogeneous helical Luttinger liquid}
\noindent Following Ref.\cite{dolcini1}, the full Hamiltonian density
\begin{equation}
\label{eq:htot}
{\cal H}(x)={\cal H}_0 (x) + {\cal H}_{{\rm int}}(x)+{\cal H}_{{\rm R}}(x)
\end{equation}
 can be diagonalized by introducing a new fermionic spinor $X(x) = (\chi_+(x), \; \chi_-(x) )^T$. It is related to the original one by the relation
\begin{equation}
	\Psi(x) = e^{\frac{i}{2}\sigma_1 \theta_{\rm R}(x)} X(x)
\end{equation}
with the Rashba angle defined by
\begin{equation}
	\theta_{\rm R}(x) = \arctan \frac{\alpha_{\rm R}(x)}{v_{\rm F}}.
\end{equation}
It is thus possible to write
\begin{equation}
\mathcal{H}_0(x) + \mathcal{H}_{\rm R}(x) = -\frac{i}{2} X^\dagger(x) v(x)\sigma_3\partial_x X(x)
\end{equation}
where
\begin{equation}
v(x) = v_{\rm F}\;  \sqrt{1+\left(\frac{\alpha_{\rm R}(x)}{v_{\rm F}}\right)^2}
\end{equation}
represents a renormalized velocity due to the Rashba coupling \cite{dolcini1}.

The density-density e-e interactions can be handled in this new basis using bosonization technique{\cite{giamarchi2003quantum, safi1995transport, calzona1, dolcini1}}, \textit{i.e.} by expressing the fermionic fields as vertex operators of two bosonic fields
\begin{equation}
	\chi_\pm = \frac{e^{i\sqrt{\pi} \left[ \Theta(x) \pm \Phi (x)\right]}}{\sqrt{2\pi a(x)}},
\end{equation}
which fulfill the commutation relation $[\Phi(x), \partial_y \Theta(y)] = 
-i \delta(x-y)$. We have introduced a space-dependent cutoff $a(x) = v(x)/E_c$ that is expressed in terms of the ultraviolet energy cutoff $E_c$ \cite{dolcini1}.
 The whole Hamiltonian can  then be expressed in the quadratic form \cite{dolcini1}
\begin{equation}
	\label{eq:H}
	\begin{split}
		H &\simeq \int \frac{ u(x)}{2} \left[K(x)\left({\partial_x \Theta(x)}\right)^2 + \frac{\left({\partial_x \Phi(x)}\right)^2}{K(x)} \right]\, dx,\\
	\end{split}
\end{equation}
where the inhomogeneous Luttinger parameter
\begin{equation}
\label{eq:Kx}
	K(x) = \sqrt{\frac{2\pi v(x) -g(x) v_{\rm F}^2 v(x)^{-2}}{2\pi v(x) + g(x)}}
\end{equation}
quantifies the {\it effective} interaction strength and
\begin{equation}
\label{eq:ux}
	u(x) =  v(x) \sqrt{\left(1-\frac{g(x) v_{\rm F}^2}{2\pi \, v(x)^3}\right)\left(1+\frac{g(x)}{2\pi \, v(x)}\right)}
\end{equation}
represents the space-dependent propagation velocity of the plasmonic modes. We stress that for free fermions $g(x)=0$, so that we get $K(x)=1$ regardless 
of $\alpha_{\rm R}$. By contrast, a finite $g(x)>0$ leads to a Luttinger parameter $0<K<1$. Interestingly, in this case, the presence of Rashba coupling $\alpha_{\rm R}>0$ further modifies the effective interaction strength. In particular it increases the value of $K$, effectively reducing e-e strength.
 
It is worth to note that we assume the spatial variations of parameters $g(x)$ and $\alpha_{\rm R}(x)$ to be smooth with respect to the Fermi wavelength, in order to neglect electronic backscattering, but abrupt with respect to wavelength of the low-energy plasmonic modes. {Consistently with 
this constraint, in obtaining Eq.\ \eqref{eq:H}, we have neglected the backscattering terms proportional to $\sin \theta_{\rm R}(x)^2 (\chi_+^\dagger \chi_- + \text{h.c.})$.{\cite{dolcini1, privitera2020}}}

\subsection{Scattering states of the inhomogeneous system}
A convenient way to address the effects of the spatial inhomogeneities of 
the Hamiltonian in Eq.\ \eqref{eq:H} is to study the scattering problem of the plasmonic modes{\cite{calzona1, muller, filippone, perfetto, degiovanni}}. For each scattering state with energy $\omega$, the equation of motion of the bosonic field $\Phi(x)$ reads
\begin{equation}
	-\omega^2 \Phi(x) = u(x)K(x) \partial_x \left[ \frac{u(x)}{K(x)} \partial_x \Phi(x) \right].
\end{equation}
In the presence of a discontinuity of the two parameters $u(x)$ and $K(x)$ at $x=\bar x$, the transmission and reflection coefficients can be determined by imposing the continuity  of $\Phi(x)$ and $u(x) K(x)^{-1}\partial_x\Phi(x)$ for $x=\bar x$. In particular, distinguishing between scattering from the left ($\rightarrow$) and from the right ($\leftarrow$), 
we find
\begin{align}
	\label{eq:r}
	r_{\rightarrow/\leftarrow} & = \mp \exp{\left[\pm \frac{2 i \omega \bar x}{ u_{L/R}}\right]} \frac{K_L-K_R}{K_L+K_R},\\
	t_{\rightarrow/\leftarrow} &= 2 \frac{K_{R/L}}{K_L+K_R} \exp \Big[i \omega \bar x \left(\frac{1}{u_L}-\frac{1}{u_R}\right)\Big],
\end{align}
where $u_L$ and $K_L$ ($u_R$ and $K_R$) are the values of velocity and Luttinger parameter to the left (right) of $x=\bar x$. The transfer matrix associated with this discontinuity point thus reads 
\begin{equation}
	T_{\bar x} = \begin{pmatrix}
		t_\rightarrow t_\leftarrow - r_\rightarrow r_\leftarrow &\,\,\,\,\,\, r_\leftarrow\\
		-r_\rightarrow & 1
	\end{pmatrix} \frac{1}{t_\leftarrow}.
\end{equation}

In the presence of $n\geq 1$ multiple inhomogeneities (spatial variations of parameters) located at $\bar x_j$ (with $\bar x_{j+1}>\bar x_j$), the transfer matrix for the whole system reads $T = T_{\bar x_n} \dots T_{\bar x_2}T_{\bar x_1}$. A straightforward computation of the corresponding scattering 
matrix gives the squared modulus of the transmission coefficient $|t(\omega)|^2$ through the whole helical region, \textit{i.e.} from one lead to the opposite one.
In App. \ref{app:tw}, we provide the full analytical expression for $|t(\omega)|^2$ in the presence of a single Rashba gate, that is for the setup depicted in Fig.\ \ref{fig:setup}.

It is important to stress that the (two terminal) electrical conductance of the helical system is given by $G_0/2=e^2/h$, regardless of the presence of inhomogeneities of $K(x)$ and $u(x)$ \cite{safi1995transport, steinberg2007charge, kamata2014fractionalized, perfetto, Tarucha1995}. The latter are known to affect the electrical transport properties only at finite frequency \cite{calzona1, kamata2014fractionalized, muller, perfetto}. By contrast, the thermal conductance does depend on the inhomogeneities of the system \cite{fazio,filippone}. This opens up the possibility to externally control the thermal transport properties of the system by selectively varying the Rashba SOC in some finite spatial regions, thus realizing a tunable thermal valve, as we will discuss in the next section.

\section{Results and discussion}
\label{sec:results}
{In this section, we will focus on thermal transport properties of the inhomogeneous helical liquid, in the presence of a thermal gradient between two non-interacting leads. We will assume that there is no voltage bias (equal chemical potentials) between them. Particular focus will be put on the effect of one (or more) top gates and how it can affect the transport behavior by inducing variation of the Rashba coupling strength.} {

\subsection{Thermal conductance}
If the two leads are kept at different temperatures $T_L$ and $T_R$, the average thermal current which flows through the helical region can be computed by integrating over all the plasmonic modes\cite{kane, fazio, filippone} 
\begin{equation}
	\label{eq:Je}
	J_E = \int_0^\infty \frac{d\omega}{2\pi} \omega |t(\omega)|^2 \left[ 
	n(\omega,T_L) - n(\omega,T_R)
	\right],
\end{equation}
where $n(\omega,T) = (e^{\omega/T}-1)^{-1}$ is the Bose distribution associated to plasmonic modes.  
Assuming a small thermal gradient between the two terminals $\Delta T = 
T_L-T_R \to 0$, \textit{i.e.} being in the linear response regime, the termal conductance reads
\begin{equation}
\label{eq:kappa}
	\kappa = \lim_{\Delta T\to 0}\frac{J_E}{\Delta T} = \int_0^\infty \frac{d \omega}{8 \pi} \frac{\omega^2}{T^2} \frac{|t(\omega)|^2}{\sinh(\omega/(2T))^2},
\end{equation}
where $T=(T_L+T_R)/2$
For a homogeneous system, in absence of e-e and Rashba interactions (\textit{i.e.} 
with $g=\alpha_{\rm R}=0$), one has $|t(\omega)|^2=1$ and thus a thermal conductance $\kappa_0$ which complies with the Wiedemann-Franz law{\cite{Benenti17}} $\kappa_0 = L T G_0/2$, where 
$L=\pi^2k_{\rm B}^2 e^{-2}/3$ is the Lorenz number\cite{Benenti17, kane, fazio}.  The presence of finite 
e-e interactions \cite{kane, fazio,krive, filippone} and, as we will show below, a finite Rashba coupling, however can strongly modify 
the behaviour of the thermal conductance and can also lead to a violation 
of the WF law. For interacting one-dimensional systems such violation has been recently observed{\cite{grossviolation}}. 

To begin the discussion, we focus on the effect of a single top gate, as shown in the geometry sketched in Fig.\ \ref{fig:setup}. The thermal conductance $\kappa$ depends then on five main parameters: the mean temperature $T$, the e-e interactions strength in the helical region $g_h$, its (possibly different) screened counterpart under the top gate $g_{{\rm R}}$, the Rashba strength $\alpha_{\rm R}$ and the covering ratio $r$ (\textit{i.e.} the 
fraction of the helical channel covered by the top gate). To facilitate comparisons with the existing literature on interacting helical liquids{\cite{calzona1, muller, strunz}}, instead of using directly the coupling constants $g_{h/{{\rm R}}}$, we hereafter denote the strength of e-e interactions in terms of the corresponding Luttinger parameters for \textit{vanishing Rashba coupling}, \textit{i.e.}
\begin{equation}
K_{h/{{\rm R}}} = \sqrt{\frac{2\pi v_{\rm F}-g_{h/{{\rm R}}}}{2\pi v_{\rm F}+g_{h/{{\rm R}}}}}.
\end{equation} 

\begin{figure}[t]
	\centering
	\includegraphics[width=.9\linewidth]{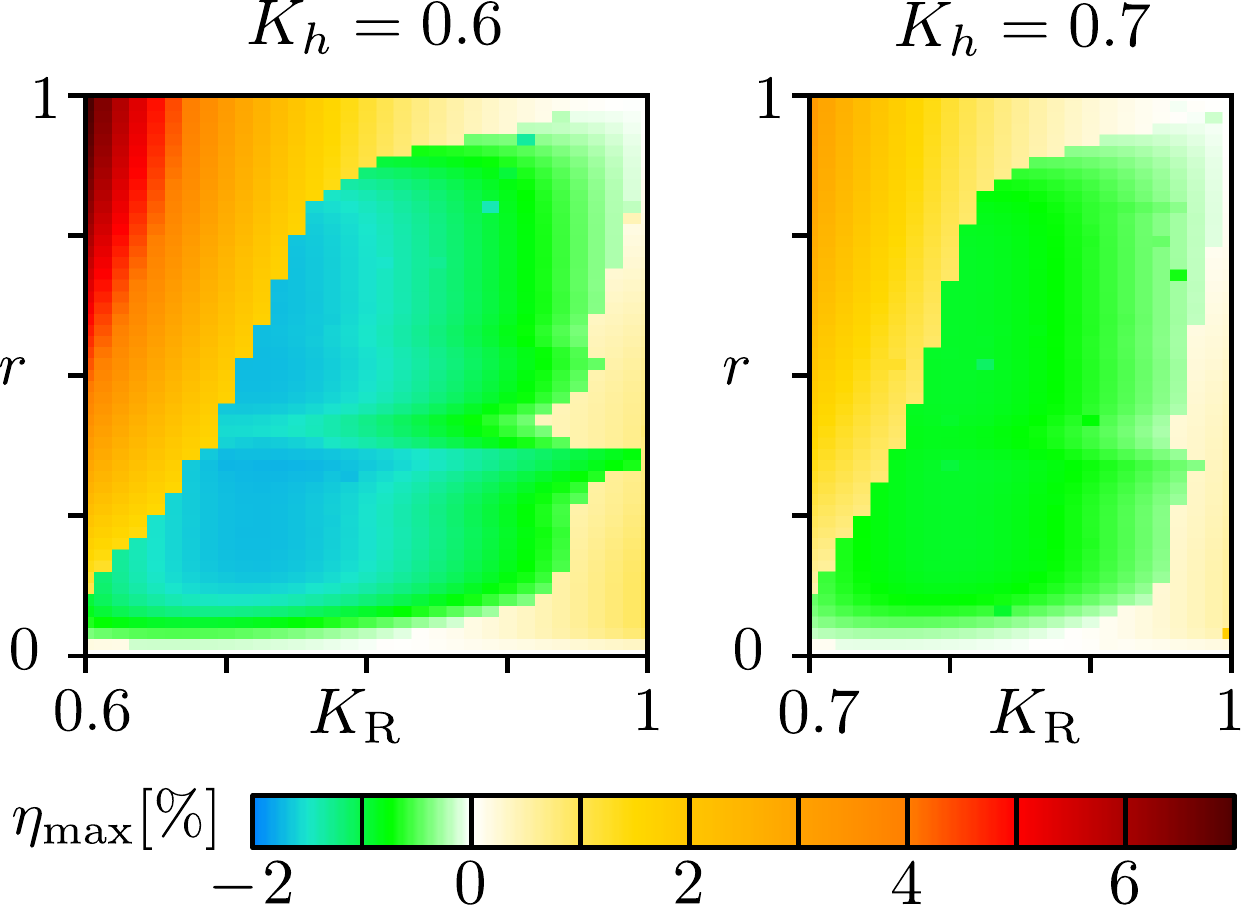}
	\caption{Plots of $\eta_{\rm max}$ (expressed as a percentage) as a function of the Luttinger parameter $K_{\rm R}$ and the size of the top gate $r$ covering the helical system. We consider a moderate interaction strength $K_h=0.6$ for the left panel and a weaker one $K_h=0.7$ in the right one. Parameters: $\alpha_{\rm max}=0.6 \, v_{\rm F}$ and $T_{\rm max} = 6\,  \epsilon$.}
	\label{fig:density}
\end{figure}

As it is clear from the above expression, a variation of the Rashba coupling can induce variations of the effective helical coupling strength which, in turn, will modulate the transmission coefficient and the resulting thermal conductance. In essence, this provides a knob to externally induce and amplify violation of the WF law. One can exploit this fact to engineer a {\it Rashba induced} thermal valve on the helical system.\\

To quantify the performance of this system as a thermal valve, we introduce the tunability ratio 
\begin{equation}
\label{eq:eta}
	\eta = \frac{\kappa(\alpha_{\rm R}=0)-\kappa(\alpha_{\rm R}=\alpha_{\rm max})}{\kappa(\alpha_{\rm R}=0)}
\end{equation}\\
which depends on the mean temperature $T$, interaction strengths, covering ratio $r$ and maximum Rashba coupling achievable $\alpha_{\rm max}$. We 
also define 
\begin{equation}
\eta_{\rm max} = \eta(\bar T)\quad \text{with }\;  |\eta(\bar T)| = \max_{0\leq T\leq T_{max}}|\eta(T)|
\end{equation}
as the tunability ratio with the highest absolute value within a fixed range of temperatures $0<T<T_{\rm max}$. In the following, we restrict the discussion to values of Rashba coupling strength in accordance with the ones that can be obtained in HgTe quantum wells\cite{Vayrynen2011}: we consider values up to $\alpha_{\rm max} = 0.6\, v_{\rm F}$. As for the temperature, we choose $T_{\max}= 6\, \epsilon$ with $\epsilon = v_{\rm F} l^{-1}$ that represents the energy scale set by the Fermi velocity and the length $l$ of the helical channels.

In Fig.\ \ref{fig:density}, we analyze the maximal tunability ratio of the system $\eta_{\rm max}$ as a function of $K_R$ and $r$, for two different values of the interaction parameter $K_h$. It is possible to identify two main interesting parameter regions. The largest tunability ratio is achieved for (i) large gate $r\sim 1$ with low screening effect $K_{\rm R}\sim K_h$, corresponding to the top left corners of the density plots in Fig.\ \ref{fig:density}. For $K_h = 0.6$ (left panel), it is possible to achieve tunability ratios around $\eta_{\rm max}\sim7\%$ 
while weaker interaction strengths lead to smaller tunabilities: for $K_h=0.7$ (right panel) one has $\eta_{\rm max}$ around $3\%$. Interestingly, in the regime (i), the sign of the tunability ratios is \textit{positive} 
(yellow-to-red colors in Fig.\ \ref{fig:density}), meaning that the onset 
of a finite Rashba coupling $\alpha_{\rm R}>0$ leads to a \textit{reduction} of the thermal conductance [see Eq.\ \eqref{eq:eta}]. Conversely, a different regime is represented by (ii) smaller covering ratios $0.2 \lesssim r \lesssim 0.8$ and moderate gate screening effect $K_{\rm R} \sim (1+K_h)/2$. In this case the maximum tunability turns out to be \textit{negative}: see the green and light blue areas located in the center of both panels in Fig.\ \ref{fig:density}. As we will carefully discuss below, in 
regime (ii) the system can indeed operate in a different way so that the presence of a finite Rashba coupling $\alpha_{\rm R}>0$ leads to an \textit{increase} of the thermal conductance. Here, the maximal tunability ratio exceeds $|\eta_{\rm max}|\gtrsim1.5\%$ for $K_h=0.6$ (left panel) and it is reduced to $|\eta_{\rm max}|\gtrsim1\%$ for weaker interactions $K_h=0.7$ (right panel). 

In passing, it is worth commenting on two (trivial) limiting behaviours. Indeed, as expected, we observe that $|\eta_{\rm max}|$ goes to $0$ both in the case of complete absence of the top gate ($r=0$) and in the presence of a top gate which covers the whole sample while completely screening the e-e interaction ($r=K_{\rm R}=1$).

\subsection{Large gate limit}

\begin{figure}[t]
	\centering
	\includegraphics[width=.9\linewidth]{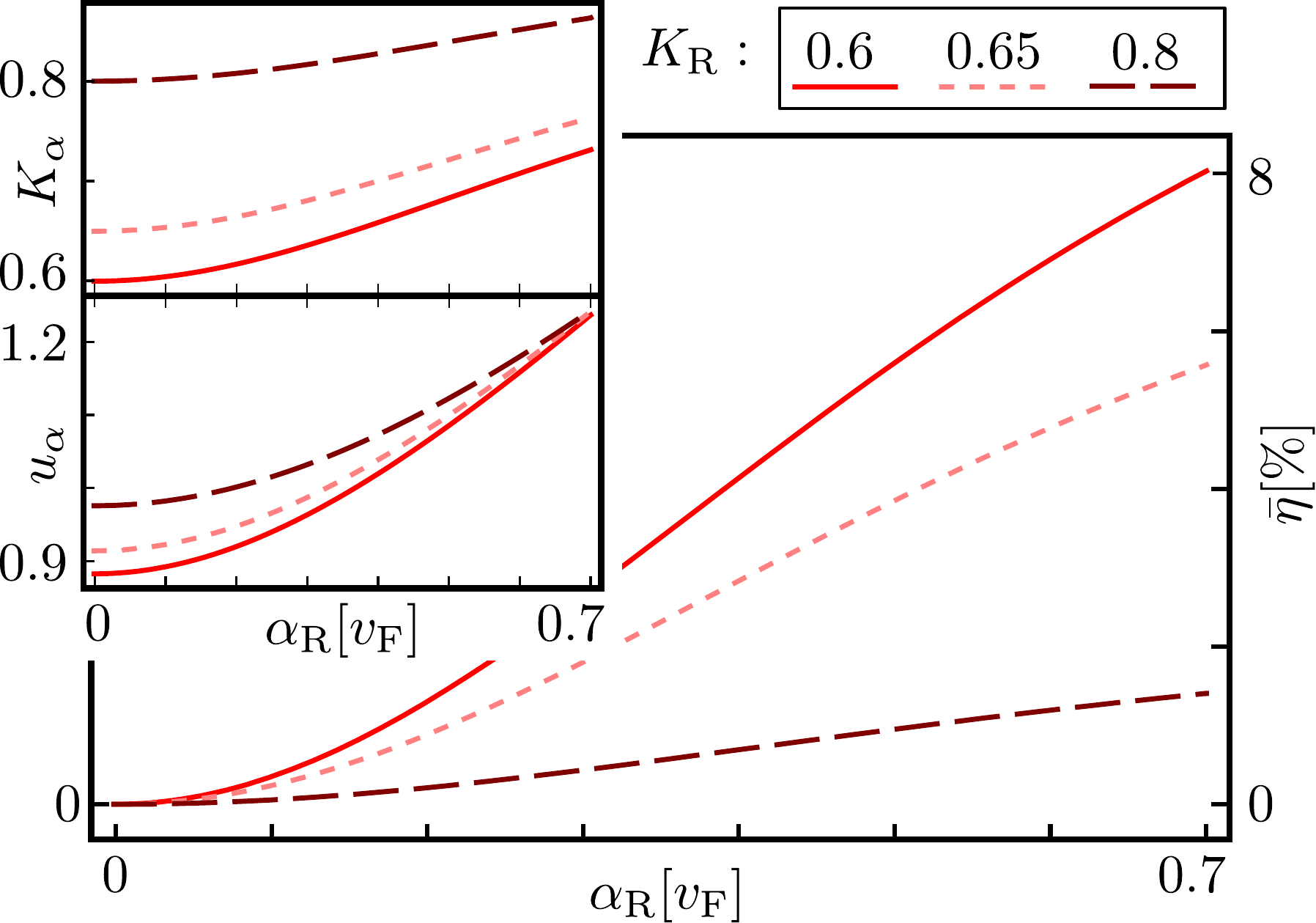}
	\caption{Plot of the tunability ratio $\bar \eta$ in the  high-temperature limit [given by Eq.\ \eqref{eq:bar_eta}] as a function of the Rashba coupling $\alpha_{\rm R}$ for a covering ratio $r=1$. The insets show the functions $K_\alpha$ and $u_\alpha$ (the latter in units $v_{\rm F}$) as a function of $\alpha_{\rm R}$.  Three different values of $K_{\rm R}=0.6, 0.65, 0.8$ are considered (see legend).
	}
	\label{fig:figKa}
\end{figure}

In order to understand the behavior of the system in the presence of a large top gate, it is useful to consider the limiting case of a covering ratio $r=1$. In this limit, the transmission probability $|t(\omega)|^2$, 
whose general analytic expression is reported in App. \ref{app:tw}, simplifies to the periodic function
\begin{equation}
\label{eq:tw}
|t(\omega)|^2 = \frac{8 K_\alpha^2}{1+6K_\alpha^2+K_\alpha^4-(1-K_\alpha^2)^2 \cos\left(2\omega l u_\alpha^{-1}\right)},
\end{equation}
where
\begin{equation}
	\label{eq:Kalpha}
	K_\alpha = \sqrt{\frac{(1+K_{\rm R}^2)(v_{\rm F}^2+\alpha_{\rm R}^2)^{3/2} - v_{\rm F}^3(1-K_{\rm R}^2)}{(1+K_{\rm R}^2)(v_{\rm F}^2+\alpha_{\rm R}^2)^{3/2} + v_{\rm F}(v_{\rm F}^2+\alpha_{\rm R}^2)(1-K_{\rm R}^2)}}
\end{equation}
and 
\begin{equation}
	\frac{u_\alpha}{v_{\rm F}} = \sqrt{\frac{\left[\left(1+\tfrac{\alpha_{{\rm R}}^2}{v_{\rm F}^2}\right)^{3/2}\!\!\!-\tfrac{1-K_{\rm R}^2}{1+K_{\rm R}^2}\right]\left[\left(1+\tfrac{\alpha_{{\rm R}}^2}{v_{\rm F}^2}\right)^{1/2}\!\!\!+\tfrac{1-K_{\rm R}^2}{1+K_{\rm R}^2}\right]}
		{1+\tfrac{\alpha_{{\rm R}}^2}{v_{\rm F}^2}}}
\end{equation}
are obtained from Eqs.~\eqref{eq:ux} and \eqref{eq:Kx}. It is important to notice that both $K_\alpha$ and $u_\alpha$ increase monotonically with $\alpha_{\rm R}^2$ (for $K_{\rm R}<1$), see the insets of Fig.\ \ref{fig:figKa}. Therefore, the presence of a finite Rashba coupling reduces both the amplitude 
and the frequency of the oscillations featured by $|t(\omega)|^2\leq 1$. The thermal conductance, defined in Eq.\ \eqref{eq:kappa}, consists of the integral of $|t(\omega)|^2$, weighted by a function which is exponentially suppressed for energies greater than the mean temperature $T$ (see Appendix \ref{app:T} for more details). Therefore, in the low temperature limit $T\to 0$, one has $\kappa=\kappa_0$, while a lower asymptotic value
\begin{equation}
	\bar\kappa = \frac{\kappa_0 l}{\pi u_\alpha} \int_0^{\tfrac{\pi u_\alpha}{l}} |t(\omega)|^2 d\omega = \frac{2K_\alpha}{1+K_\alpha^2} \, \kappa_0 
\end{equation}
is reached for $T\gtrsim \epsilon$. The corresponding tunability ratio thus becomes
\begin{equation}
	\label{eq:bar_eta}
	\bar \eta = \frac{K_\alpha}{K_{\rm R}}\frac{1+K_{\rm R}^2}{1+K_\alpha^2}-1,
\end{equation}
which is plotted in Fig.\ \ref{fig:figKa}. To better understand the magnitude of this tunability ratio $\bar \eta$, note that with interaction parameter $K_{\rm R} = 0.6$ and $\alpha_{{\rm R}}=\alpha_{\rm max} (= 0.6 v_{\rm F})$, we get $\bar \eta \simeq 6.9\%$ . 
At low temperature ($T\lesssim \epsilon$), the above considerations qualitatively hold also for a slightly smaller covering ratio. This is shown in Fig.\ \ref{fig:figlo}, where it is numerically characterized a system with $r=0.95$ and a moderate interaction strength. Fig. \ref{fig:figlo}(a) shows the thermal conductance as a function of temperature,  for different values of $\alpha_R$, with $K_h=0.6$ and a small screening effect leading to $K_{\rm R}=0.65$. The inset displays 
the corresponding transmission coefficients $|t(\omega)|^2$. The latter feature the expected oscillating pattern [see Eq.\ \eqref{eq:tw}], while the thermal conductance decreases from $\kappa=\kappa_0$ at $T=0$, to values close to $\bar \kappa$ (see the horizontal grey lines) for $T\sim \epsilon$. Fig. \ref{fig:figlo}(b)  displays the tunability ratio $\eta$ for different choices of the interaction parameters $K_h$ and $K_{\rm R}$. At low temperature, there is almost no dependence on $K_h$. The highest tunability ratios are achieved for $T\sim 0.5\, \epsilon$ and they are well described by the analytical result $\bar \eta$ in Eq.\ \eqref{eq:bar_eta} (see the horizontal grey lines). For completeness, in the inset of Fig.\ \ref{fig:figlo}(b), it is shown how the tunability ratios at fixed temperature $T=0.5\, \epsilon$ depend on the choice of $\alpha_{\rm max}$: as expected, the higher the Rashba coupling, the higher the tunability of the system.

\begin{figure}[t]
	\centering
	\includegraphics[width=.9\linewidth]{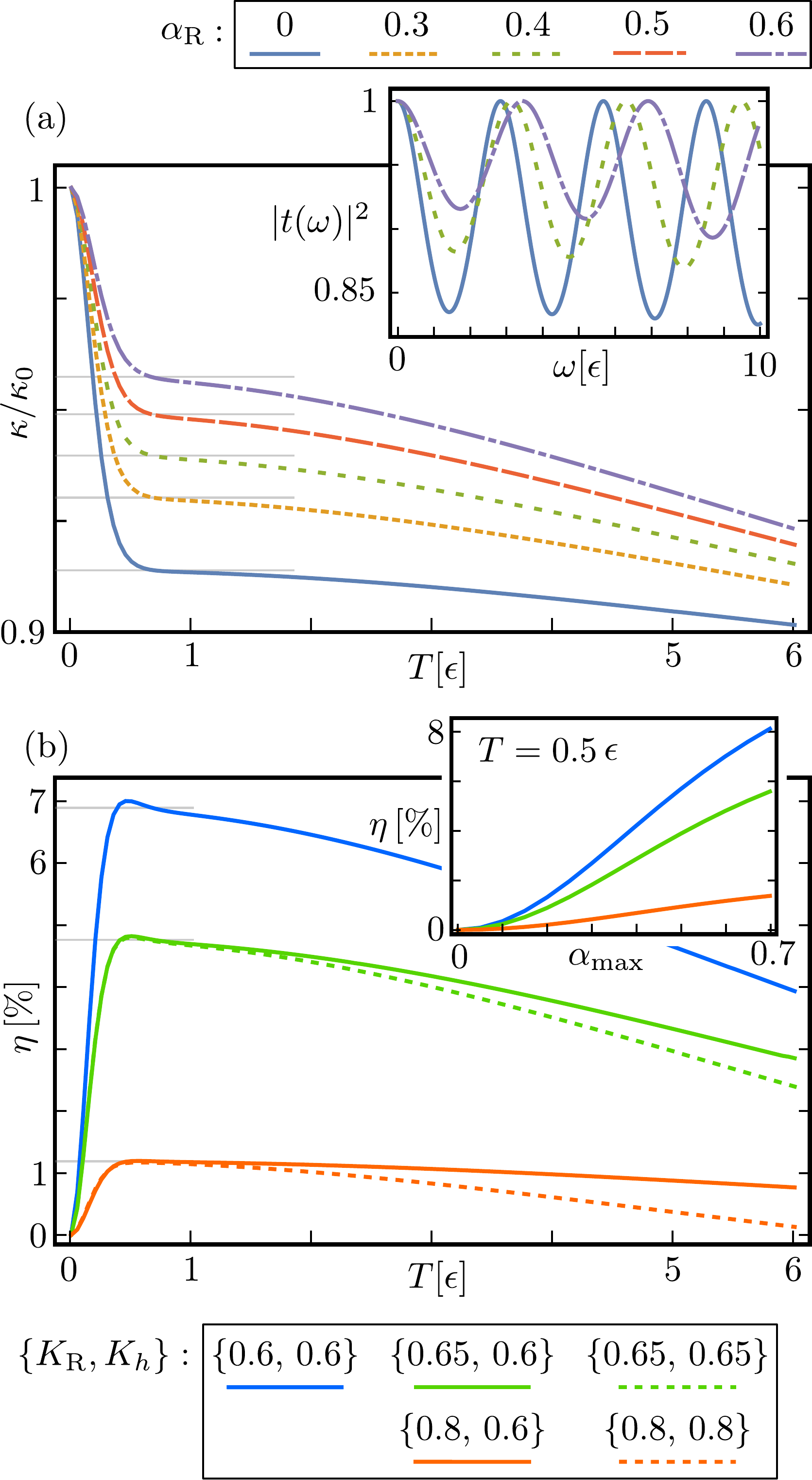}
	\caption{Characterization of the setup with a large Rashba gate $r=0.95$. Panel (a) displays the thermal conductance $\kappa$ as a function of the temperature $T$, for five different values of the Rashba coupling $\alpha_{{\rm R}}$ (see legend). The horizontal reference lines in grey shows 
		$\bar \kappa/\kappa_0$ for the considered values of $\alpha_{{\rm R}}$. The inset displays the transmission coefficient $|t(\omega)|^2$ for three values of the Rashba coupling (see legend). Other parameters are: $K_h = 
		0.6$ and $K_{{\rm R}} = 0.65$. Panel (b) displays the tunability ratio $\eta$ (with $\alpha_{\rm max} = 0.6 v_{\rm F}$) as a function of the temperature $T$ and for five different choices of parameters $K_{{\rm R}}$ 
		and $K_h$ (see legend). The horizontal reference lines in grey shows the $\bar \eta$ for the three different choices of $K_{\rm R}$. The inset shows the dependence of $\eta$ on $\alpha_{\rm max}$ (units $[v_{\rm F}]$) for a fixed temperature $T=0.8\, \epsilon$. 
	}
	\label{fig:figlo}
\end{figure}

The presence of short helical regions not covered by the central gate for 
$r\lesssim 1$ induces interference effects in the propagation of the plasmonic modes with energies of the order of $\omega \gtrsim \epsilon (1-r)^{-1}$ or higher. In particular, destructive interferences are responsible for the high-temperature reduction of the thermal conductance below $\bar \kappa$ displayed in Fig.\ \ref{fig:figlo}(a). The effects of the 
short helical regions emerges also in Fig.\ \ref{fig:figlo}(b), where tunability ratios show a visible dependence on $K_h$ for temperatures above 
$T \gtrsim \epsilon$. However, it is important to stress that for large gate regime, the most interesting temperature range, \textit{i.e.} where the largest tunability can be achieved, is the low-temperature one, where the effects of the short helical regions are negligible. The situation is quite different for values of the covering ratio $r$ significantly smaller than $1$. In what follows, we will then focus on the short gate regime and analyze the interference effects more in detail.
\begin{figure}[t]
	\centering
	\includegraphics[width=.9\linewidth]{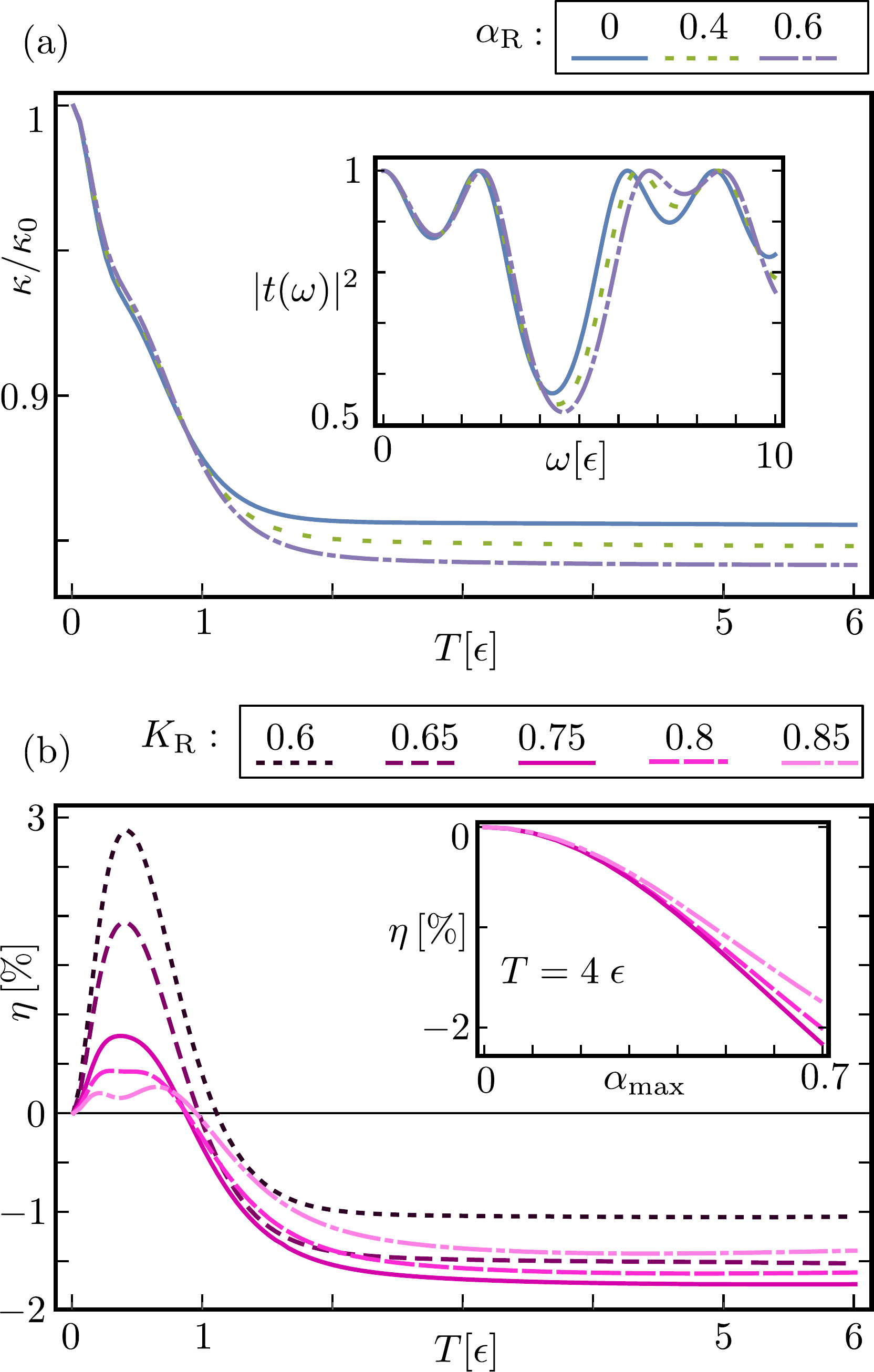}
	\caption{Characterization of the setup with a short Rashba gate $r=1/3$. Panel (a) displays the thermal conductance $\kappa$ as a function of the temperature $T$, for three different values of the Rashba coupling $\alpha_{{\rm R}}$ (see legend). The inset displays the corresponding transmission coefficient $|t(\omega)|^2$. Other parameters are: $K_h = 0.6$ and $K_{{\rm R}} = 0.8$. Panel (b) displays the tunability ratio $\eta$ (with $\alpha_{\rm max} = 0.6 v_{\rm F}$) as a function of the temperature $T$ and for $K_h=0.6$ and five different values of $K_{{\rm R}}$ (see legend) corresponding to different screening effects of the top gate. The inset shows the dependence of $\eta$ on $\alpha_{\rm max}$ (units $[v_{\rm F}]$) for a fixed temperature $T=4\, \epsilon$ and three choices of $K_{\rm R}$ (see legend).}
	\label{fig:figsh}
\end{figure}

\subsection{Short Rashba gate}
\label{sec:short}
In Fig.\ \ref{fig:figsh}(a), we numerically analyze the properties of the setup with a short gate ($r=1/3$), moderate interaction strength $K_h=0.6$ and moderate gate screening which leads to $K_{\rm R}=0.8$. In this case, the energy scales 
associated with the plasmonic scattering under the top gate and in the remaining helical regions are of the same order of magnitude $\sim 3\epsilon$. Interference effects play therefore an important role in shaping the transmission coefficient $|t(\omega)|^2$, which is carefully analyzed in Appendix \ref{app:tw} and plotted in the inset of Fig.\ \ref{fig:figsh}(a). It features a first local minimum around $\omega \sim \epsilon$, which is reminiscent of the minima appearing in the long gate limit [see the inset of Fig.\ref{fig:figlo} (a)]. At slightly higher energies, 
however, destructive interference effects lead to a significantly deeper minimum around $\omega \sim 4 \epsilon$. The effect of a finite Rashba coupling on these two minima is different: while finite values of $\alpha_{\rm R}$ barely affect the first minimum (and make it less pronounced), the depth of the second one is increased. The features of $|t(\omega)|^2$ directly affect the thermal conductance, which is plotted in the main panel of Fig.\ \ref{fig:figsh}(a). For temperatures of the order of $0.5 \, \epsilon$, only the first minimum of the transmission coefficient is relevant. As a result, the dependence of the thermal conductance on the Rashba coupling is weak and only a strong Rashba coupling leads to a slightly higher $\kappa$. At higher temperatures, however, the second and 
more pronounced minimum of $|t(\omega)|^2$ starts playing a more central role. This leads to a significant reduction of $\kappa$ around $0.85 \, \kappa_0$ and to an opposite dependence of the thermal conductance on the Rashba strength: higher values of $\alpha_{\rm R}$ decrease the value of $\kappa$.\\

This behavior also emerges in the tunability ratio $\eta$ (with $\alpha_{\rm max} = 0.6 v_{\rm F}$), which is plotted in Fig.\ \ref{fig:figsh}(b) as a function of $T$. Here, we set $K_h=0.6$ and consider different values for $K_{\rm R}$.
 In general, the tunability ratio is positive for low temperature and drops below zero for higher temperatures, eventually becoming constant for $T\gtrsim 3 \, \epsilon$. In this regime, $\eta$ shows a  non-monotonic dependence on $K_{\rm R}$: the maximal tunability is indeed achieved 
for $K_{\rm R} \sim 0.75$ while both weaker and stronger values are detrimental for the performance of the system as a thermal valve. This can be understood as follows. The destructive interference effects, which affect the transmission coefficient at high energy, are enhanced whenever the reflection coefficients for each inhomogeneity are stronger. According to Eq.\ \eqref{eq:r}, this requires a large difference between the 
Luttinger parameters $K_h$ and $K_\alpha$. Given the dependence of the latter on $\alpha_{\rm R}$ and $K_{\rm R}$ [see Eq.\ \ref{eq:Kalpha}], it is clear that the tunability of the system is maximized when (i) $K_{\rm R}$ is sufficiently smaller than $1$ so that $K_\alpha$ has a stronger dependence on $\alpha_{\rm R}$ and (ii) $K_{\rm R}$ is sufficiently greater than $K_h$ so that interference effects can be relevant. For the sake of completeness, in the inset of Fig.\ \ref{fig:figsh}(b), we also show the dependence of $\eta$ on the choice of $\alpha_{\rm max}$ at a fixed value 
of $T=4\, \epsilon$.

To summarize, in the presence of a short top gate, interference effects already dominate for $T\simeq 3 \epsilon$. In this regime, the tunability of the thermal valve is maximized for a moderate screening effect of the gate, \textit{i.e.} for a $K_{\rm R}$ which differs significantly from both $K_h$ and $1$. As an example, for $K_h=0.6$ and $K_{\rm R}=0.75$ $(0.8)$ the solid (long-dashed) line of Fig.\ \ref{fig:figsh}(b) shows that it is possible to achieve $\eta < -1.5\%$. This value is smaller than the tunability ratios achievable in the regime (i), \textit{i.e.} with a \textit{large gate} and \textit{negligible screening of the top gate}. However, it is important to stress that, in the presence of \textit{moderate screening} (e.g. $K_h=0.6$ and $K_{\rm R}=0.8$), the tunability is actually maximized by exploiting the interference effects associated to a \textit{short gate} at temperatures $T\gtrsim 3 \, \epsilon$: see Fig.\ \ref{fig:density} and compare the solid orange line in Fig.\ \ref{fig:figlo}(b) with the long-dashed line in Fig.\ \ref{fig:figsh}(b). 

\begin{figure}[h!]
	\centering
	\includegraphics[width=.85\linewidth]{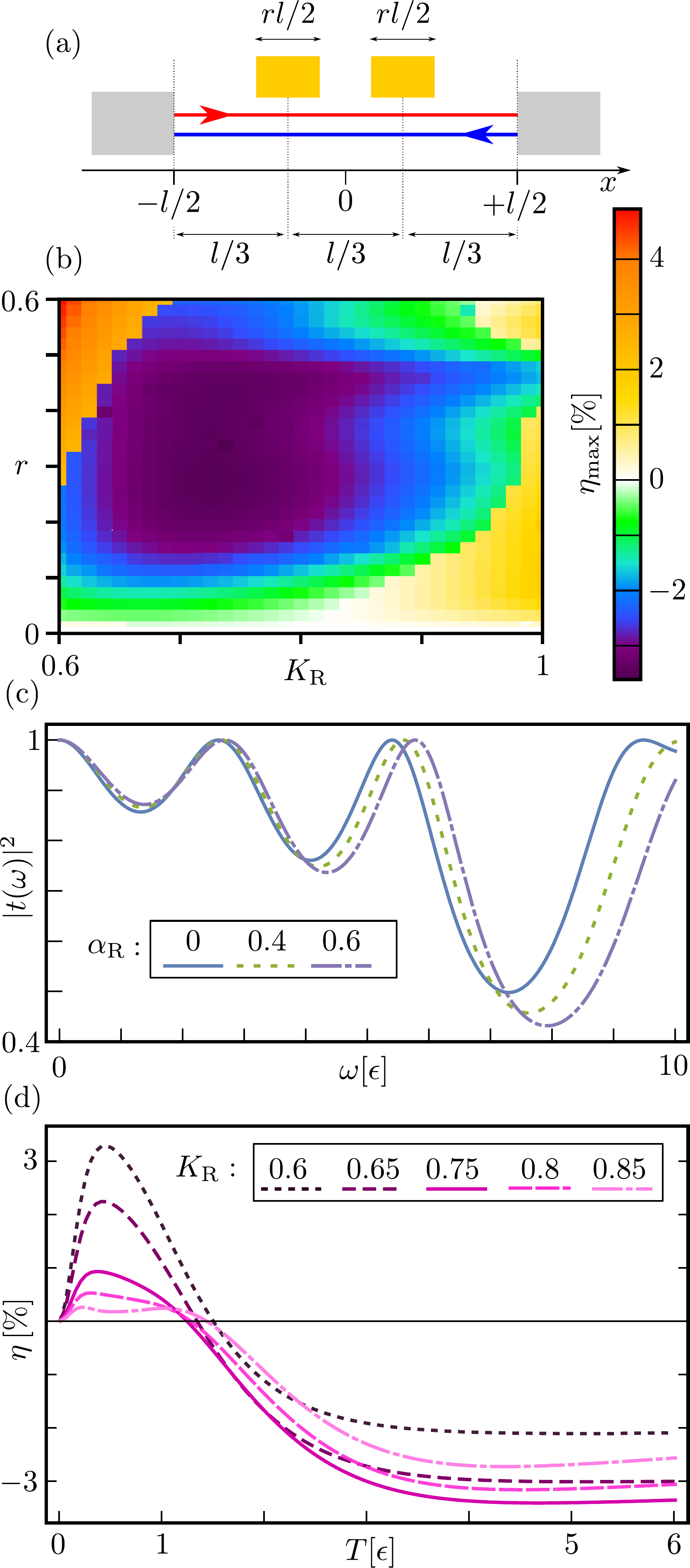}
	\caption{Characterization of the setup with two short Rashba gates. Panel (a): sketch of the setup, with two identical gates (yellow) centered around $x=\pm l/6$, each one covering a fraction $r/2$ of the helical region (red and blue lines) in between the two leads (in grey). Panel (b): plot of $\eta_{\rm max}$ (expressed as a percentage) for $K_h=0.6$ as a function of the Luttinger parameter $K_{\rm R}$ and the total fraction of 
the system which is covered a gate $r$ (with $T_{\rm max} = 6 \epsilon$ and $\alpha_{\rm max}=0.6\, v_{\rm F}$). Panel (c): transmission coefficient $|t(\omega)|^2$ for three values of the Rashba coupling (see legend) for $K_h=0.6$, $K_{\rm R}=0.75$ and $r=0.4$. Panel (d): tunability ratio $\eta$ (with $\alpha_{\rm max}=0.6\,v_{\rm F}$) as a function of the temperature $T$ for $r=4, $$K_h=0.6$ and five different 
values of $K_{\rm R}$ (see legend) corresponding to different screening effects of the gates. 
	}
	\label{fig:double}
\end{figure}

\subsection{Double gate geometry}
In the previous section, we have shown the importance of interference effects that arise in the presence of three regions with different Luttinger 
parameters and comparable length. It is therefore natural to ask whether these effects could be boosted by adding other inhomogeneities to the system. To answer this question, we now consider a different setup, sketched in Fig.\ \ref{fig:double} (a). For the sake of simplicity, it consists of two identical short top gates centered around $x=\pm l/6$, each one covering a fraction $r/2$ of the region 
between the leads. A different positioning of the gates and differences in their length and/or bias would only affect the results at a quantitative level. 

At first, along the lines of the previous sections, we plot $\eta_{\rm max}$ as a function of $K_{\rm R}$ and $r$ for a fixed moderate interaction 
strength in the helical regions $K_h=0.6$. The resulting density plot in Fig.\ \ref{fig:double}(b) is qualitatively similar to its single gate counterpart [see Fig.\ \ref{fig:density}(a)]. However, the parameter regime (ii) shows values of $\eta_{\rm max}$ well below $-3\%$. With respect to the single gate configuration, in regime (ii) the transmission coefficient $|t(\omega)|^2$ features an additional third deep minimum, located at 
$T\sim 7\, \epsilon$ and highly dependent on the Rashba coupling. 
This is shown in Fig.\ \ref{fig:double}(c) for $K_h=0.6$, $K_{\rm R} = 
0.75$ and a covering ratio $r=0.4$. As shown in Fig.\ \ref{fig:double}(d), for temperatures around $T\sim 4\, \epsilon$, this allows to achieve negative tunability ratio as low as $-3.5\%$. This would result in a doubling of the performance of the system as a thermal valve in the moderate-screening regime.

\subsection{Indication of helicity}
The underlying principle behind the results just obtained basically rely
on the fact that the thermal conductance depends on the spatial
variations of the Luttinger parameter. Such variations can, in any
one-dimensional interacting fermionic system, be induced by gates, that
locally screen the electronic interactions. A second mechanism is
however present in helical liquids. Indeed, in this case, the gates also
modify the Rashba spin-orbit coupling, which in turn renormalizes the
velocity of propagating modes. Since the Luttinger parameter, just like
the Wigner radius in usual Fermi liquids, gives a measure of the
relative importance of the kinetic and the potential energy, the spatial
variations of the Rashba coupling do induce an extra contribution to the
thermal conductance. Interestingly, an experiment disentangling the
effect of screening from the effect of Rashba coupling could help
answering the question: Are the edge states of two-dimensional
topological insulators spin polarized or helical? Indeed, in the case of
spin polarized edges, no Rashba-induced renormalization of the propagation velocity and of the Luttinger parameter can
happen. This means that if one finds a Rashba contribution to the
thermal conductance, then the channel is helical. While it is hard to
conceive an experiment, involving gating, able to disentangle the two
effects, a possibility is provided by the geometric manipulation of the
Rashba interaction by means of a proper shaping of the
edges\cite{ortix}. In this case, samples with different shape should be
characterized by the same strength of the electron-electron
interactions, but, if helical, by different profiles of the Rashba
coupling. By comparing the thermal conductance in samples with different
shape, one could hence determine if the edges are helical or spin
polarized. On the other hand, the case of edge states with four degrees
of freedom per edge can be simply discriminated by inspecting the
conductance quantization.

\section{Summary and conclusions}
\label{sec:summary}
In this work, we have considered the role of the Rashba coupling in a helical system, in presence of Coulomb interactions. The presence of a capacitively coupled top gate allows to tune its strength, inducing also a partial screening of electronic interactions felt by the helical liquid. Based on an inhomogeneous Luttinger liquid approach, we have considered different regions with spatially varying Rashba coupling and interaction strength, by solving the associated EOMs with a scattering approach of plasmonic modes. These results have been applied to characterize the thermal conductance of the helical channels in presence of a thermal gradient in a two-terminal configuration. Provided that the e-e interaction strength is finite, we have shown that it is possible to control the value of the thermal conductance by properly tuning the Rashba strength. The system can therefore act as a gate-controlled thermal valve. Its associated performances have been characterized in the linear response regime (small thermal gradient) in a wide parameter region, considering both large and short top gates.

It is shown that the larger tunability of the thermal valve can be achieved for (mean) temperature of the order of the energy scale $\epsilon = \hbar v_{\rm F}l^{-1}\, $ ( characteristic energy scale set by the length of the system $l$ and the Fermi velocity $v_{\rm F}$) or higher. Moreover, the performance of the system increases with the e-e interaction strength and/or with the gate-induced Rashba coupling.

 It is argued that, if the screening effect of the top gate is small, so that the helical system beneath the gate still feels moderate Coulomb interactions, a large top gate configuration (covering almost the whole helical channels) can act as an efficient thermal valve. Considering for example an interaction strength $K_{\rm R} \sim 0.6$, the thermal conductance of the system can be \textit{increased} by $7\%$ by ramping up the gate-induced Rashba SOC from $\alpha_{{\rm R}} = 0 \to 0.6 v_{\rm F}$. However, the performance of such a system is dramatically reduced in presence of a stronger screening effect of the top gate. In this case, provided that the helical regions not covered by the gate still feel moderate interactions, it is convenient to consider alternative setups with a shorter gate. In this case, the increase of the Rashba SOC $\alpha_{{\rm R}} = 0 \to 0.6 v_{\rm F}$ can lead to a \textit{reduction} of the thermal conductance slightly smaller than $2\%$ (for $K_h=0.6$ and $K_R=0.75$). This figure can be almost doubled by adding a second short gate alongside the first one. Indeed, the presence of additional spatial inhomogeneities results in more complicated interference patterns (with deeper minima in the transmission coefficients) that eventually are responsible for the obtained increase of the figure of merit. 

A brief comment on the experimental feasibility can be made by estimating the associated energy scales discussed above 
using state-of-the-art numbers for system hosting helical channels. In HgTe, one has $v_{\rm F} = 5\, 10^5\, m/s$ and it is reasonable to consider $l=4\mu m$. This leads to $\epsilon \simeq 1 K$. 
Moreover, quoting Ref. \cite{Vayrynen2011}, ``in case the gate voltage between the well and the gate electrode 100 nm apart is 1 V, the estimated value of the Rashba coupling is of the order of $\alpha \sim 0.16 nm eV = 
0.44 \hbar v_{\rm F}$". Of course, several other materials showing helical 
channels and SOC interactions, like semiconducting nanowires{\cite{Du15}}, Bismuthene flakes {\cite{stuhler20}}, or engineered high-quality graphene devices {\cite{veyrat}} can be exploited as alternative platforms which can eventually achieve more interesting parameter regimes. Finally, we mention that all the qualitative analyses presented here should still hold even away from the linear response regime. Indeed, for larger temperature gradients, the function which weights the transmission coefficient would be exponentially suppressed with the higher of the two temperatures as shown in Appendix \ref{app:T}.

\begin{acknowledgments}
We acknowledge Bj\"orn Trauzettel and Sankalp Gaur for early stage contributions to this work.
\end{acknowledgments}
\appendix
\section{The transmission coefficient}
\label{app:tw}
In this Appendix, we provide the analytical expression of the transmission coefficient $|t(\omega)|^2$ in the presence of a single gate which covers a fraction $r$ of the helical channels, as sketched in Fig.\ \ref{fig:setup}. In total, we thus have four inhomogeneities located at $x=\pm l$ and $x=\pm rl$. By multiplying the corresponding transmission matrices and computing the resulting scattering matrix, we obtain the analytical 
expression 
\begin{equation}
	\label{eq:tCII}
	|t(\omega)|^2 = 128 K_h^4 K_{\alpha}^2 \Omega^{-1}
\end{equation}
with the denominator
\begin{widetext}
\begin{equation}
	\begin{split}
		\Omega & = 6 K_h^8 + 4 K_h^6 (1+K_{\alpha}^2) + 2 K_h^4 (3+44K_\alpha^2+3K_\alpha^4)+4K_h^2 K_\alpha^2 (1+K_\alpha^2) + 6 K_\alpha^4\\
		&+ 2(K_h^2-1)^2(K_h^2-K_\alpha^2)^2 \cos\left[2\omega l u_h^{-1}  (1-r) 
\right] \\
		&+ 4(1-K_h^4)(K_{\alpha}-K_h)(K_\alpha+K_h)  (K_h+K_\alpha)^2 \cos \left[\omega l u_h^{-1} \left(2 r \frac{u_h}{u_\alpha}+1-r\right)\right] \\
		&+ 4(1-K_h^4)(K_\alpha-K_h)(K_\alpha+K_h)(K_h-K_\alpha)^2 \cos \left[\omega l u_h^{-1} \left(2 r \frac{u_h}{u_\alpha}+r -1  \right)\right]\\
		&-(1-K_h^2)^2 (K_h+K_\alpha)^4 \cos \left[2 \omega l u_h^{-1} \left(r\frac{u_h}{u_\alpha}-r+1 \right)\right]\\
		&-(1-K_h^2)^2 (K_h-K_\alpha)^4 \cos \left[2 \omega l u_h^{-1} \left(r\frac{u_h}{u_\alpha}+r-1 \right)\right]\\
		&-2 (3+2K_h^2+3K_h^4)(K_h^2-K_\alpha^2)^2 \cos\left[2r  \omega l u_{\alpha}^{-1}\right]\\
		&-8 (1-K_h^4)(K_\alpha^4-K_h^4) \cos\left[\omega l u_h^{-1} (1-r) \right].
	\end{split}
\end{equation}
\end{widetext}
For $r=1$, the expression for $|t(\omega)|^2$ reduces to Eq.\ \eqref{eq:tw} and the denominator features only one single cosine function. By contrast, in the most generic case ($r<1$ and $K_h \neq K_\alpha$) the transmission coefficient contains seven different cosines which interfere with 
each other. This behavior is evident in the inset of Fig.\ \ref{fig:figsh}(a), where we plot $|t(\omega)|^2$ for $r=1/3$ and highlight the presence of different minima.    

\begin{figure}
	\centering
	\includegraphics[width=0.8\linewidth]{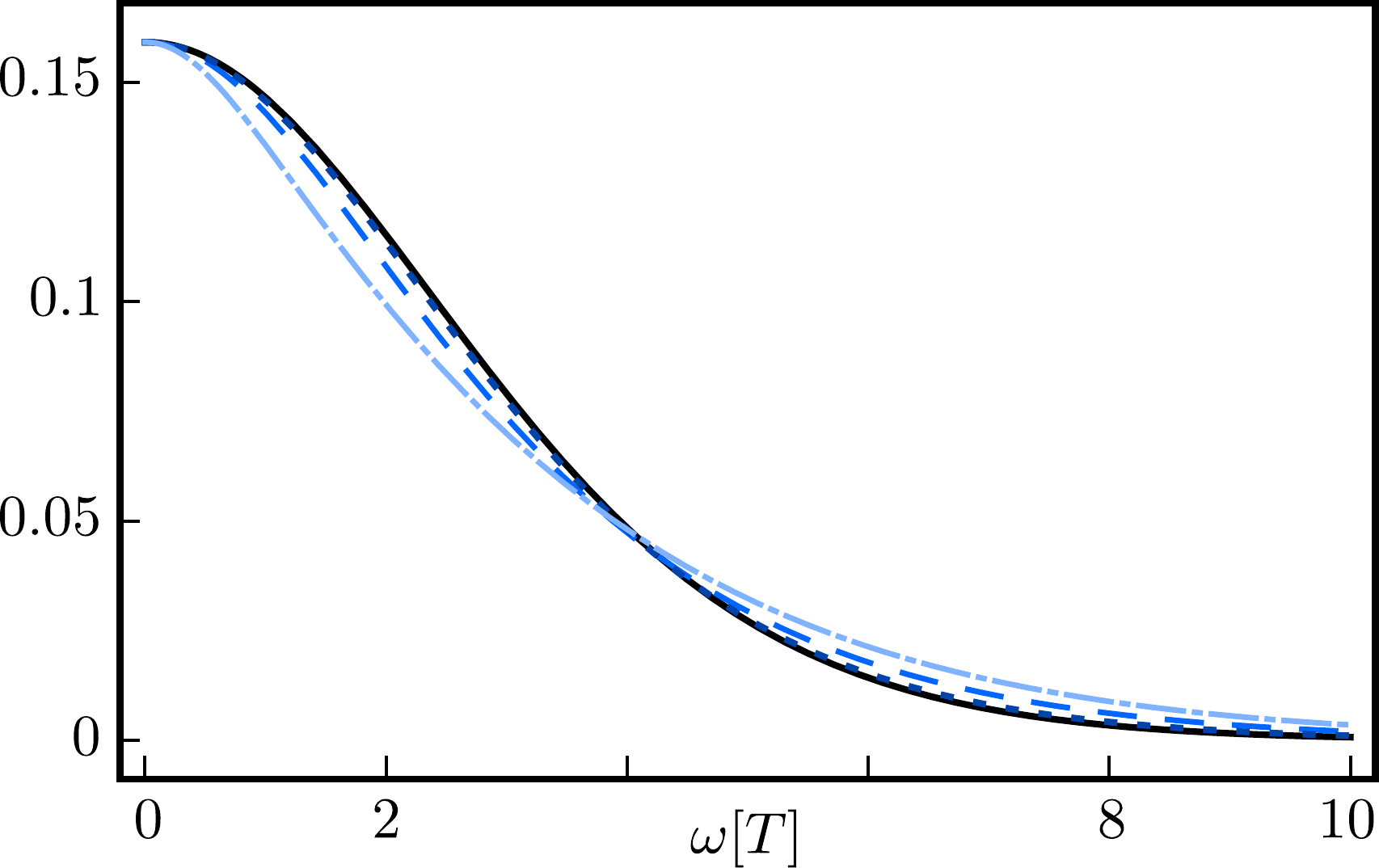}
	\caption{The solid black line shows the weight function $W_\kappa$ as a 
function of $\omega$ (units $T$). The other lines show $W_J/(T_h-T_c)$ for different choices of $T_c$ and $T_h$ obeying $T_h+T_c=2T$. In particular $T_h=1.25T$ for the dark blue dotted line, $T_h=1.5T$ for the blue dashed line and $T_h=1.75T$ for the light blue dotted-dashed line.}
	\label{fig:figapp}
\end{figure}

\section{Non linear regime}
\label{app:T}
The thermal conductance $\kappa$ is defined in the linear regime, \textit{i.e.} for a small thermal gradient $\Delta T$ between the two leads [see Eq.\ \eqref{eq:kappa}]. Such a regime, however, might be difficult to study in experiments since the measure of small energy current $J_E$ is extremely challenging. In this respect, from an experimental point of view, it might be easier to consider a large temperature difference between the two leads in order to enhance $J_E$. 

Importantly, the behavior of $J_E$ away from the linear regime still shares many of the qualitative features of the conductance that we analyzed in the main text. Indeed, the only difference between Eqs.\ \eqref{eq:Je} and \eqref{eq:kappa} is the function which weights the transmission coefficient within the integral over the energy. In particular, we have
\begin{align}
	J_E &= \int_0^\infty |t(\omega)|^2 W_J(\omega) d\omega\\
	\kappa &= \int_0^\infty |t(\omega)|^2 W_\kappa(\omega) d\omega
\end{align}
with
\begin{align}
	W_J &= \frac{\omega}{2\pi} \left(\frac{1}{e^{\omega/T_h}+1}-\frac{1}{e^{\omega/T_c}+1}\right)\\
	W_\kappa &= \frac{\omega^2}{8\pi T^2} \frac{1}{\sinh(\omega/2T)^2}
\end{align}
where we considered the left lead to be hotter (at temperature $T_h$). As 
long as $T_h + T_c \sim 2T$, the weight function $W_\kappa$ shares the same main features with $W_J/(T_h-T_c)$, see Fig.\ \ref{fig:figapp}.


\end{document}